# Large deformation and instability of soft hollow cylinder with surface effects

Jian Wu[a, b, d], Mingchao Liu[a, c], Zhenyu Wang[a, b], C.Q. Chen[a, c]

[a]AML, Department of Engineering Mechanics, Tsinghua University, Beijing 100084, China
[b]Center for Mechanics and Materials, Tsinghua University, Beijing 100084, China
[c]Center for Nano and Micro Mechanics (CNMM), Tsinghua University, Beijing 100084, China
[d]Center for Flexible Electronics Technology, Tsinghua University, Beijing 100084, China

## Graphic abstract

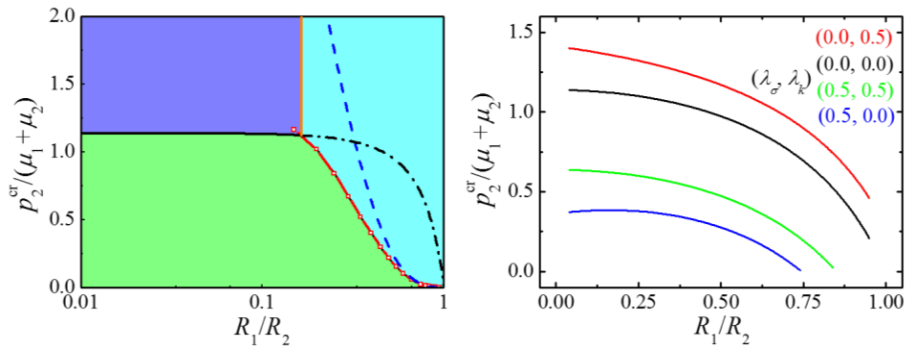

## Abstract

Surface stress, which is always neglected in classical elastic theories, has recently emerged as a key role in the mechanics of highly deformable soft solids. In this paper, the effect of surface stress on the deformation and instability of soft hollow cylinder are analyzed. By incorporating surface energy density function into the constitutive model of a hyper-elastic theory, explicit solutions are obtained for the deformation of soft hollow cylinder under the conditions of uniform pressure loading and geometric everting. It is found that surface tension evidently alters the deformation of the soft cylinder. Specifically, the surface stiffness resists the deformation, but the residual surface stress is inclined to larger deformation. Effects of surface stress on the instability of the soft hollow cylinder is also explored. For both the pressure loading and geometric everting conditions, significant changes in critical condition of the creases are found by varying the surface parameter. The results in this work reveal that surface energy obviously influences both the deformation and the instability of soft hollow cylinder at finite deformation. The obtained results will be helpful for understanding and predicting the mechanical behavior of soft structures accurately.

# 1. Introduction

Soft materials, such as tissues, elastomers and hydrogels, are widely available in nature and modern industry. They can sustain very large deformation and show great potential for engineering applications including biological system, artificial devices and soft robotics.[1-3] With the development of material synthesis technologies, many researchers have successfully designed and fabricated a variety of soft materials with excellent performance, such as high toughness, high strength and high electric conductivity.[4-6] A mutual understanding of the mechanical behavior of soft materials is of fundamental importance in fulfilling their functional applications and has received an increasing interest in recent years.[7,8]

Traditionally, the influence of surface energy is neglected in the engineering analysis of the deformation of solids.[9] However, some of the previous studies have indicated that surface energy can significantly affect the deformation of soft materials or nanostructures made of stiff materials,[10,11] and/or even be the dominant force to drive the shape change of soft materials.[12,13] To identify the relative importance of surface stress and bulk elasticity on material behavior by driving or resisting deformation, a characteristic material length scale can be approximately estimated by the ratio of surface energy density to elastic modulus.[14-16] For stiff materials, such as metals and ceramics, this length scale is in the order of a few nanometers.[17,18] However, for such soft materials as tissues, gels and polymers, the length scale increases greatly to hundreds of nanometers and even a few microns.[10,16,19] Therefore, surface energy plays important roles on the mechanical response, not only deformations but also instabilities, of soft materials.[15,20]

The concept of surface effect in solids was first introduced by Gibbs[21] and has been steadily developed since then.[22-24] Recently, the continuum elastic theory has been extended by incorporating surface energy effects to explain the unusual experimental observations for soft solids, such as edge softening,[12, 13] unexpected indentation and contact responses,[25-27] and stiffening solids with liquid inclusions.[19, 28] In most of these studies, to simplify the analysis, surface stress is assumed to be isotropic and independent of surface strain, and hence the additional displacements due to the surface stress can be linearly superimposed on the classical elastic solution. However, this simplification is only valid for the small strains. The deformation of soft solids is large, hence finite deformation models are required to correctly capture the physics and mechanics of surface stress. The pioneering work for the finite surface elasticity accounting for large deformation has been proposed by Gurtin and Murdoch.[23] More recently, the typical Gibbs surface elasticity[21] has also been extended into a large deformation regime, and used for describing the surface effects on the large deformation of soft materials.[29]

In particular, as a simple component, soft hollow cylinder is widely found in a wide range from biophysical to artificial applications, such as injection into subcutaneous tissues, blood flow through arteries and vascular networks, and design of robotics like artificial systems.[30-32] For the soft hollow cylinders under operating environments, large deformation will be generated by the pressure applied on the outer and/or inner surfaces.[33] In some of these cases, the soft cylinder is prone to instability when the load exceed a certain critical value.[34-36] Except for the external loading, the eversion, which means turning a structure inside out, is also ubiquitous for soft hollow cylinder in different fabrication processes.[37-39] The everted tube maintains smooth surface when the thickness is small, otherwise, if the thickness is larger than a critical value, the non-uniform deformation will be caused by mechanical instability.[40]

It is important to systematically understand the deformation and instability of soft hollow cylinder under loading and everting conditions. The major focus of most previous works has been limited to the elastic response and instability of soft cylinder under external mechanical loads or everting deformation. However, the surface effect on the elastic behavior of soft cylinders has attracted much less attention, even if the influence is significant for most of the soft materials.[29] In this work, we investigate the surface effect on the deformation and instability of soft hollow cylinder on the basis of finite surface elasticity. The theoretical framework considering the surface effect at finite deformation is obtained for both the external loading and everting conditions.

The structure of this paper is as follows. We summarize the constitutive relation of large deformation of the isotropic elasticity and the basic theory of surface elasticity in Section 2. The deformations of the soft hollow cylinder under the conditions of uniform pressure loading and geometric everting are determined in Section 3. Additionally, the instability and crease of the deformed hollow cylinders are investigated in Section 4. The summary and conclusions are given in Section 5.

## 2. The constitutive model for large deformation

### 2.1 The constitutive model of the large deformation in elasticity

The strain energy density of the elasticity, $W$, which dependents on the deformation gradient, $\mathbf{F}$, is often assumed to be the function of the first, second and third invariants of the right Cauchy-Green deformation tensor, $\mathbf{C}$, for isotropic materials. The three invariants of the right Cauchy-Green deformation tensor can be given by the principal stretches, $\lambda_{\mathrm{I}}$, $\lambda_{\mathrm{II}}$, $\lambda_{\mathrm{III}}$, as $\mathrm{I} = \lambda_{\mathrm{I}}^2 + \lambda_{\mathrm{II}}^2 + \lambda_{\mathrm{III}}^2$, $\mathrm{II} = \lambda_{\mathrm{I}}^2\lambda_{\mathrm{II}}^2 + \lambda_{\mathrm{I}}^2\lambda_{\mathrm{III}}^2 + \lambda_{\mathrm{II}}^2\lambda_{\mathrm{III}}^2$, $\mathrm{III} = \lambda_{\mathrm{I}}^2\lambda_{\mathrm{II}}^2\lambda_{\mathrm{III}}^2$. The third invariant is connected with the deformation ratio of volume, $J$, as $\mathrm{III} = J^2$, which is $\mathrm{III} = J^2 = 1$ for the incompressible materials.

The second Piola-Kirchhoff stress, $\mathbf{T}$, for the incompressible materials can be given as

$$\mathbf{T} = -p\mathbf{C}^{-1} + 2\left[\left(\frac{\partial W}{\partial \mathrm{I}} + \mathrm{I}\frac{\partial W}{\partial \mathrm{II}}\right)\mathbf{1} - \frac{\partial W}{\partial \mathrm{II}}\mathbf{C}\right]. \tag{1}$$

where $p$ is the unknown pressure, $\mathbf{C}^{-1}$ is the inverse of the right Cauchy-Green deformation tensor, $\mathbf{C}$. By substituting equation (1) into the relation between the Cauchy stress, $\boldsymbol{\sigma}$, and the second Piola-Kirchhoff stress, $\boldsymbol{\sigma} = J^{-1}\mathbf{F}\cdot\mathbf{T}\cdot\mathbf{F}^{\mathrm{T}}$, the Cauchy stress for incompressible materials is

$$\boldsymbol{\sigma} = -p\mathbf{1} + 2\left(\frac{\partial W}{\partial \mathrm{I}}\mathbf{B} - \frac{\partial W}{\partial \mathrm{II}}\mathbf{B}^{-1}\right), \tag{2}$$

where $\mathbf{F}$ and $\mathbf{F}^{\mathrm{T}}$ are the deformation gradient and its transpose, respectively, $\mathbf{B}$ is the left Cauchy-Green deformation tensor, $\mathbf{B} = \mathbf{F}\cdot\mathbf{F}^{\mathrm{T}}$.

The strain energy density for the Mooney-Rivlin materials[41] is

$$W = \frac{\mu_1}{2}(I-3) + \frac{\mu_2}{2}(II-3) \;, \tag{3}$$

where $\mu_1$ and $\mu_2$ are the material parameters. The constitutive relation of the Moony-Rivlin materials is

$$\boldsymbol{\sigma} = -p\mathbf{1} + \left(\mu_1 \mathbf{B} - \mu_2 \mathbf{B}^{-1}\right) \;. \tag{4}$$

**2.2 The constitutive model with surface effect in elasticity**

The surface energy, $\gamma$, on elastic body is assumed to be dependent on the surface deformation gradient, $\mathbf{F}_s$. For isotropic materials, the surface energy only depends on the first and second invariants of the surface left Cauchy-Green deformation tensor, $\mathbf{B}_s$, which can be expressed by the surface deformation gradient as $\mathbf{B}_s = \mathbf{F}_s \cdot \mathbf{F}_s^{\mathrm{T}}$. The first and second invariants of $\mathbf{B}_s$ are $I_s = \mathrm{trace}(\mathbf{B}_s)$ and $II_s = J_s^2 = \det(\mathbf{B}_s)$, where $J_s$ is the ratio of the deformed surface area to the initial surface area. The surface stress, $\boldsymbol{\sigma}_s$, is defined by the surface energy as[29, 42]

$$\boldsymbol{\sigma}_s = 2\frac{\partial \gamma}{\partial I_s}\mathbf{B}_s + \left(J_s \frac{\partial \gamma}{\partial J_s} + \gamma\right)\mathbf{1}_s \tag{5}$$

where $\mathbf{1}_s$ is the unit tensor on the deformed surface. The surface stress, $\boldsymbol{\sigma}_s$, satisfies the following equation on the surface,

$$\nabla_s \cdot \boldsymbol{\sigma}_s + \mathbf{t}_s - \boldsymbol{\sigma} \cdot \mathbf{n} = 0 \;, \tag{6}$$

where $\nabla_s$ is the gradient operator on the deformed surface, $\mathbf{t}_s$ is the external traction per unit area on the deformed surface, $\mathbf{n}$ is the unit vector along the normal direction of the deformed surface.

For simplicity, the surface energy density is assumed as[38]

$$\gamma = \sigma_0 + \frac{k_s}{2J_s}(J_s - 1)^2, \quad (7)$$

where $\sigma_0$ is the residual surface stress, and $k_s$ is the surface stiffness. By substituting equation (7) into equation (5) the surface stress is

$$\sigma_s = \left[\sigma_0 + k_s(J_s - 1)\right]\mathbf{1}_s. \quad (8)$$

## 3. Large deformation of hollow cylinder

Here we focus on the effects of surface stress on deformation of a soft hollow cylinder. Two typical loading conditions, i.e., uniform pressure and geometric everting, are considered, as shown in Fig. 1. By applying the constitutive model established in Section 2.2, explicit solutions of stress and deformation fields of the soft cylinder are obtained under both conditions.

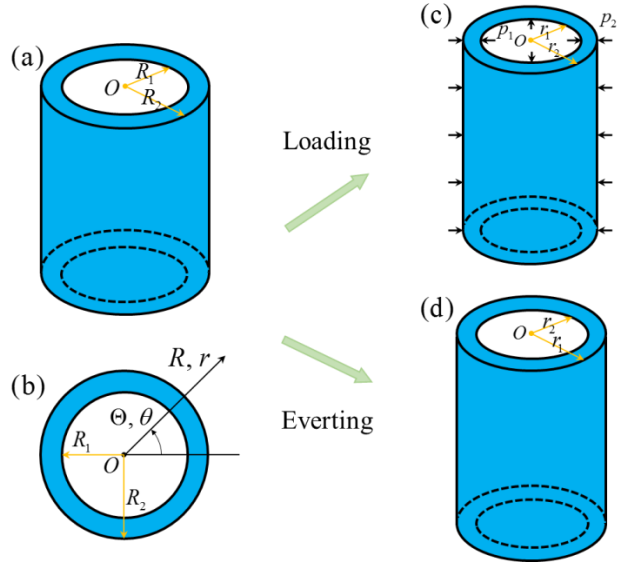

Fig. 1 Schematic of the reference configuration of a soft hollow cylinder: (a) oblique view, and (b) plan view with the coordinate system. Schematic of the deformed

configuration of the cylinder under the conditions of: (c) uniform pressure, and (d) geometric everting.

## 3.1 The large deformation of hollow cylinder under pressures

As shown in Fig. 1(a), the inner, outer radii and length of the hollow cylinder are $R_1$, $R_2$ and $L$, respectively. The coordinates of initial and deformed hollow cylinders are $(R, \Theta, Z)$ and $(r, \theta, z)$ as shown in Fig. 1(b), where the base vectors of the Cartesian coordinates of initial and deformed hollow cylinders are $(\mathbf{e}_R, \mathbf{e}_\Theta, \mathbf{e}_Z)$ and $(\mathbf{e}_r, \mathbf{e}_\theta, \mathbf{e}_z)$, respectively.

The axisymmetric deformation of the hollow cylinder under pressure (see Fig. 1(c)) can be expressed as

$$r = r(R),\ \theta = \Theta,\ z = \lambda_z Z\ , \tag{9}$$

where $\lambda_z$ is the elongation along axial direction of cylinder, the inner and outer radii $R_1$, $R_2$ of cylinder move to $r_1$, $r_2$, i.e., $r_1 = r_1(R_1)$, $r_2 = r_2(R_2)$ and $r_1 < r_2$.

The deformation gradient of hollow cylinder is $\mathbf{F} = \dfrac{dr}{dR}\mathbf{e}_r\mathbf{e}_R + \dfrac{r}{R}\mathbf{e}_\theta\mathbf{e}_\Theta + \lambda_z\mathbf{e}_z\mathbf{e}_Z$, and the left Cauchy-Green deformation tensor is $\mathbf{B} = \left(\dfrac{dr}{dR}\right)^2 \mathbf{e}_r\mathbf{e}_r + \left(\dfrac{r}{R}\right)^2 \mathbf{e}_\theta\mathbf{e}_\theta + \lambda_z^2 \mathbf{e}_z\mathbf{e}_z$. The three invariants of the left Cauchy-Green deformation tensor, which are same with the invariants of the right Cauchy-Green deformation tensor, are

$$\text{I} = \left(\frac{dr}{dR}\right)^2 + \left(\frac{r}{R}\right)^2 + \lambda_z^2,$$

$$\text{II} = \left(\frac{r}{R}\right)^2\left(\frac{dr}{dR}\right)^2 + \lambda_z^2\left(\frac{r}{R}\right)^2 + \lambda_z^2\left(\frac{dr}{dR}\right)^2, \qquad (10)$$

$$\text{III} = \lambda_z^2\left(\frac{r}{R}\right)^2\left(\frac{dr}{dR}\right)^2.$$

The third invariant, III, equals to $J^2 = 1$ due to incompressibility of materials, and the relation between the radius of the deformed and initial cylinders are given as

$$\lambda_z r\, dr = R\, dR,\ \lambda_z r^2 = R^2 + A\ , \qquad (11)$$

where $A$ is a constant.

The nonzero components of the Cauchy stress, $\sigma$, in the constitutive relation equation(2), are

$$\sigma_{rr} = -p + \left[\mu_1\left(\frac{R}{\lambda_z r}\right)^2 - \mu_2\left(\frac{\lambda_z r}{R}\right)^2\right],$$

$$\sigma_{\theta\theta} = -p + \left[\mu_1\left(\frac{r}{R}\right)^2 - \mu_2\left(\frac{R}{r}\right)^2\right], \qquad (12)$$

$$\sigma_{zz} = -p + \left(\mu_1\lambda_z^2 - \mu_2\lambda_z^{-2}\right).$$

where $\sigma_{rr}$, $\sigma_{\theta\theta}$ and $\sigma_{zz}$ are the components of the Cauchy stress in Cartesian coordinates of cylinder.

The Cauchy stress satisfy the equilibrium equations, i.e., $\nabla \cdot \sigma = 0$, where $\nabla = \frac{\partial}{\partial r}\mathbf{e}_r + \frac{1}{r}\frac{\partial}{\partial \theta}\mathbf{e}_\theta + \frac{\partial}{\partial z}\mathbf{e}_z$ is the gradient operator. By substituting the constitutive equation (12) into the equilibrium equations, the component of the Cauchy stress, $\sigma_{rr}$, satisfies the following equation

$$\frac{d\sigma_{rr}}{dr} = \left(\mu_1 + \mu_2 \lambda_z^2\right) \frac{A}{\lambda_z^2 r^3} \frac{2\lambda_z r^2 - A}{\lambda_z r^2 - A}. \tag{13}$$

For the axial-symmetric deformation of cylinder, which also satisfy the equation (11), the surface deformation gradient is $\mathbf{F}_s = \frac{r}{R} \mathbf{e}_\theta \mathbf{e}_\Theta + \lambda_z \mathbf{e}_z \mathbf{e}_Z$, and the surface left Cauchy-Green deformation tensor is $\mathbf{B}_s = \mathbf{F}_s \cdot \mathbf{F}_s^T = \left(\frac{r}{R}\right)^2 \mathbf{e}_\theta \mathbf{e}_\theta + \lambda_z^2 \mathbf{e}_z \mathbf{e}_z$, the invariants of which are

$$I_s = \left(\frac{r}{R}\right)^2 + \lambda_z^2, \quad II_s = \left(\frac{r}{R}\right)^2 \lambda_z^2 = J_s^2. \tag{14}$$

By substituting equation (14) into equation (8), the surface stress is

$$\boldsymbol{\sigma}_s = \left[\sigma_0 + k_s \left(\frac{\lambda_z r}{R} - 1\right)\right] \mathbf{1}_s. \tag{15}$$

The pressures $p_1$ and $p_2$ are applied on the inner and outer surfaces, as shown in Fig. 1(c), which satisfy the condition of equation (6), and the component of the Cauchy stress, $\sigma_{rr}$, at the inner and outer surfaces has the following boundary conditions,

$$\sigma_{rr}\big|_{R=R_1} = -p_1 + \frac{\sigma_0}{r_1} + k_s \left(\frac{\lambda_z}{R_1} - \frac{1}{r_1}\right), \tag{16}$$

$$\sigma_{rr}\big|_{R=R_2} = -p_2 - \frac{\sigma_0}{r_2} - k_s \left(\frac{\lambda_z}{R_2} - \frac{1}{r_2}\right). \tag{17}$$

By integrating equation (13), the stress $\sigma_{rr}$ which satisfies the boundary condition of equation (16) can be obtained as

$$\sigma_{rr} = -p_1 + \frac{\sigma_0}{r_1} + k_s \left(\frac{\lambda_z}{R_1} - \frac{1}{r_1}\right) + \left(\mu_1 + \mu_2 \lambda_z^2\right) \frac{1}{2\lambda_z^2} \left[\frac{R^2}{r^2} - \frac{R_1^2}{r_1^2} + \lambda_z \ln\left(\frac{R^2}{R_1^2} \frac{r_1^2}{r^2}\right)\right]. \tag{18}$$

The other nonzero components of the Cauchy stress are given in Appendix.

The stress $\sigma_{rr}$ also satisfies the boundary condition of equation (17), which drives the relation between the deformation and the pressures under non-elongation along axial direction (i.e., $\lambda_z = 1$),

$$\frac{p_2 - p_1}{\mu_1 + \mu_2} = -\lambda_k - \frac{\lambda_\sigma - \lambda_k}{1 + \lambda_R} \frac{1}{\lambda_{\theta 1}} \left\{ 1 + \left[ 1 + \left( \lambda_R^{-2} - 1 \right) \lambda_{\theta 1}^{-2} \right]^{-\frac{1}{2}} \right\}$$
$$+ \frac{1}{2} \frac{1}{\lambda_{\theta 1}^2} \left\{ 1 - \lambda_R^{-2} \left[ 1 + \left( \lambda_R^{-2} - 1 \right) \lambda_{\theta 1}^{-2} \right]^{-1} \right\} + \frac{1}{2} \ln \left\{ \lambda_R^2 \left[ 1 + \left( \lambda_R^{-2} - 1 \right) \lambda_{\theta 1}^{-2} \right] \right\}$$
(19)

where $\lambda_R = \dfrac{R_1}{R_2}$, $\lambda_{\theta 1} = \dfrac{r_1}{R_1}$, $\lambda_\sigma = \dfrac{\sigma_0}{\mu_1 + \mu_2} \left( \dfrac{1}{R_1} + \dfrac{1}{R_2} \right)$, $\lambda_k = \dfrac{k_s}{\mu_1 + \mu_2} \left( \dfrac{1}{R_1} + \dfrac{1}{R_2} \right)$.

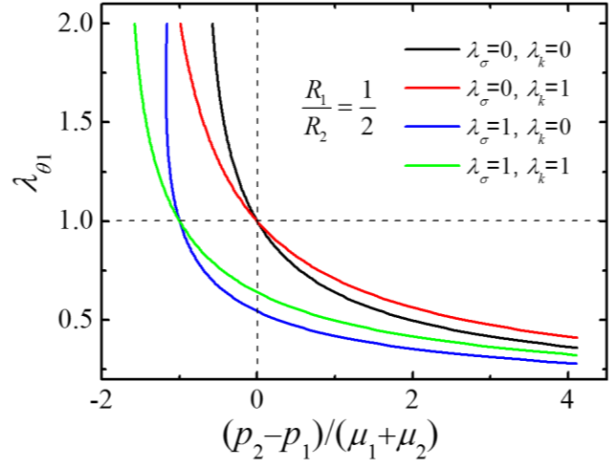

Fig. 2 The circumferential deformation of the soft hollow cylinder versus normalized pressure load for different values of surface parameters at a fix initial radius ratio $R_1/R_2=1/2$.

Figure 2 shows the normalized residual surface stress and normalized surface stiffness have significant effects on the deformation of the pressured hollow cylinder. The normalized surface stiffness, $\lambda_k$, resists the deformation. The positive surface stiffness increases the stiffness of the whole cylinder, which mean larger pressure or tensile load is required to achieve the same deformation of the hollow cylinder without surface effect. The effect of the normalized residual surface stress, $\lambda_\sigma$, on the deformation of cylinder depends on the deformation state of cylinder. For the positive residual surface stress, the smaller pressure or larger tensile load are required to achieve a given deformation.

The stretches along circumferential and radial directions of the cylinder without elongation along axial direction are $\lambda_\theta = \dfrac{r}{R}, \ \lambda_r = \dfrac{R}{r}$. The stretch distributions satisfy the following equations

$$\frac{\lambda_\theta^2 - 1}{\lambda_{\theta 1}^2 - 1} = \frac{R_1^2}{R^2}, \ \frac{\lambda_r^{-2} - 1}{\lambda_{r1}^{-2} - 1} = \frac{R_1^2}{R^2}, \tag{20}$$

where $\lambda_{r1}$ is the stretch along the radial direction at inner surface.

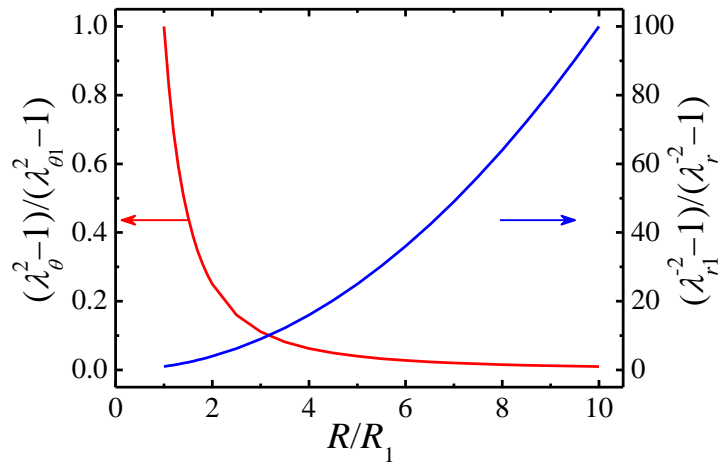

Fig. 3 The distributions of circumferential and radial deformations along the radial position of the soft hollow cylinder under the uniform pressure loading condition.

The distributions of stretch are plotted in Fig. 3. In the circumferential direction, the compressive/tensile deformation decrease from maximum value at the inner surface to minimum value at the outer surface. In the radial direction, the maximum deformation also appears at inner surface due to the incompressibility of material, but there are the tensile/compressive deformations. The surface stiffness and the residual surface stress have no effect on the distribution modes of the cylinder deformation, but the magnitudes of the cylinder deformation depend on the pressures and the surface parameters.

**3.2 The large deformation of everted hollow cylinder**

The soft hollow cylinder can be everted by turning the inner surface out,[37-40] as shown in Fig. 1(d). The deformation of the everted hollow cylinder is similar to the pressurized one, but $\lambda < 0$ in equations (9), (11), and (13) for everted cylinder. For long hollow cylinder, the elongation along the axial direction of the everted cylinder can be neglected, i.e., $\lambda_z = -1$.

The Cauchy stress in an everted cylinder also satisfy the constitutive equation (12) and the equilibrium equation, $\nabla \cdot \boldsymbol{\sigma} = 0$. The component of Cauchy stress, $\sigma_{rr}$, at the initial inner and outer surface of everted cylinder without pressures are

$$\sigma_{rr}\big|_{R=R_1} = -\frac{\sigma_0}{r_1} - k_s\left(\frac{1}{R_1} - \frac{1}{r_1}\right), \tag{21}$$

$$\sigma_{rr}\big|_{R=R_2} = \frac{\sigma_0}{r_2} + k_s\left(\frac{1}{R_2} - \frac{1}{r_2}\right). \tag{22}$$

The stress, $\sigma_{rr}$, which satisfies the boundary condition (22) and the equation (13), is

$$\sigma_{rr} = \frac{\sigma_0}{r_2} + k_s \left( \frac{1}{R_2} - \frac{1}{r_2} \right) + (\mu_1 + \mu_2) \frac{1}{2} \left[ \left( \frac{R^2}{r^2} - \frac{R_2^2}{r_2^2} \right) - \ln \left( \frac{R^2}{R_2^2} \frac{r_2^2}{r^2} \right) \right]. \qquad (23)$$

The other nonzero components of the Cauchy stress are given in Appendix.

The stress, $\sigma_{rr}$, in equation (23) also satisfies the boundary condition (21), which gives the relation between the deformation, geometry and the surface parameters of the cylinder as

$$\frac{\lambda_\sigma - \lambda_k}{1 + \lambda_R^{-1}} \frac{1}{\lambda_{\theta2}} \left[ \frac{1}{\sqrt{1 + (1 - \lambda_R^2) \lambda_{\theta2}^{-2}}} + 1 \right] + \lambda_k$$

$$+ \frac{1}{\lambda_{\theta2}^2} \left[ \frac{\lambda_R^2}{1 + (1 - \lambda_R^2) \lambda_{\theta2}^{-2}} - 1 \right] - \ln \left[ \frac{\lambda_R^2}{1 + (1 - \lambda_R^2) \lambda_{\theta2}^{-2}} \right] = 0 \qquad (24)$$

where $\lambda_{\theta2} = r_2 / R_2$.

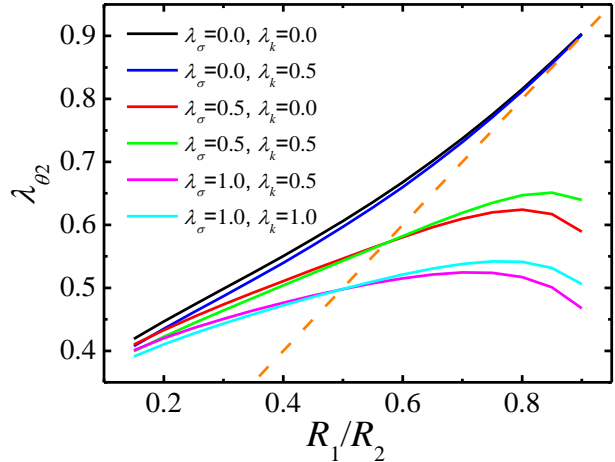

Fig. 4 The circumferential deformation of the soft hollow cylinder versus initial radius ratio for different values of surface parameters.

Figure 4 shows the surface effect on the deformation of the everted cylinder. The normalized residual surface stress, $\lambda_\sigma$, has significant effect on the deformation of the everted cylinder, which is inclined to reduce the stiffness of the cylinder. The effect of the normalized surface stiffness, $\lambda_k$, depends on the geometry and deformation of the cylinder. There is a critical line in Figure 4, the normalized surface stiffness, $\lambda_k$, resists the deformation on one side below the critical line, and raises the deformation on the side above the critical line. The critical line for the deformation at initial outer surface, $\lambda_{\theta 2}$, is

$$\left[(1+\lambda_R)\lambda_{\theta 2} - \lambda_R\right]\sqrt{\lambda_{\theta 2}^{~2} + 1 - \lambda_R^{~2}} - \lambda_R \lambda_{\theta 2} = 0 , \qquad (25)$$

which is also given as orange dash line in Figure 4.

The distributions of the principal stretches in the everted cylinder satisfy the following equations,

$$\frac{\lambda_\theta^2 + 1}{\lambda_{\theta 2}^2 + 1} = \frac{R_2^2}{R^2}, \quad \frac{\lambda_r^{-2} + 1}{\lambda_{r2}^{-2} + 1} = \frac{R_2^2}{R^2} , \qquad (26)$$

where $\lambda_{r2} = R_2/r_2$.

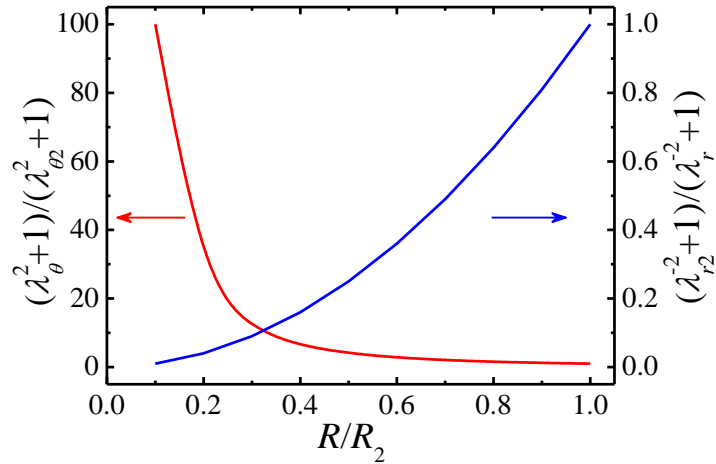

Fig. 5 The distributions of circumferential and radial deformations along the radial position of the soft hollow cylinder under the geometric everting condition.

The stretch distributions of everted cylinder are shown in Fig. 5. In the circumferential direction, the cylinder is compressive at initial outer surface ($R = R_2$), and tensile at initial inner surface ($R = R_1$), there is a neutral circle, where no deformations happen, on the cross section of the cylinder. The radius of neutral circle is $R_0 = R_2\sqrt{(\lambda_{\theta 2}^2 + 1)/2}$, which depends on the geometry and surface parameters of the hollow cylinder. In the cross section of the everted cylinder, the compressive deformation increases from zero at the neutral circle to maximum value at the initial out surface, and the tensile deformation increases from zero at the neutral circle to the maximum value at the initial inner surface.

## 4. The instability of hollow cylinder

The surface of the elasticity is instable under the large compressive state, where the creases will appear when surface is compressed to the critical strain. Hong et al.[44] studied the surface crease of soft matter and gave the corresponding critical condition under generalized plane-strain state as

$$\lambda_\perp / \lambda_\parallel = \beta , \qquad (27)$$

where $\lambda_\parallel$ is the stretch along the compressive direction in the surface, $\lambda_\perp$ is the stretch along the direction perpendicular to the compressive direction, and $\beta = 2.4$ is for crease.

The both surfaces of the compressed hollow cylinder (i.e., $p_2 - p_1 > 0$) are compressive along the circumferential direction, and the initial outer surface is compressed along the circumferential direction of the everted cylinder. The maximum compressive deformations along circumferential direction appear at the initial inner and outer surfaces for the pressured and everted cylinders, respectively, and $\lambda_\parallel = \lambda_\theta$, $\lambda_\perp = \lambda_r$ for the critical strain, which give the critical condition for creases of the pressured cylinder is

$$\left(\frac{\lambda_r}{\lambda_\theta}\right)_{R=R_1} = \beta . \tag{28}$$

By substituting equation (28) into equation (19), the critical pressure, $p_2^{cr}$, where the pressure is only applied on the outer surface, for the crease is

$$\begin{aligned}\frac{p_2^{cr}}{\mu_1 + \mu_2} = &-\frac{\lambda_\sigma - \lambda_k}{\lambda_R + 1}\sqrt{\beta}\left\{1 + \left[1 - \beta\left(1 - \lambda_R^{-2}\right)\right]^{-\frac{1}{2}}\right\} - \lambda_k \\ &+ \frac{\beta}{2}\left\{1 - \lambda_R^{-2}\left[1 - \beta\left(1 - \lambda_R^{-2}\right)\right]^{-1}\right\} + \frac{1}{2}\ln\left\{\lambda_R^{2}\left[1 - \beta\left(1 - \lambda_R^{-2}\right)\right]\right\}\end{aligned} . \tag{29}$$

For the thin cylinder, the pressure also causes the buckling behavior of the whole cylinder, not only crease on surface, the critical pressure for the buckling of the cylinder is[55]

$$\frac{p_2^{cr}}{\mu_1 + \mu_2} = \left[2\left(1 - \frac{R_1}{R_2}\right)\bigg/\left(1 + \frac{R_1}{R_2}\right)\right]^3 . \tag{30}$$

Figure 6 shows the normalized critical pressure for different surface parameters. The normalized surface stiffness, $\lambda_k$, resists forming the creases in surface, and the normalized residual surface stress, $\lambda_\sigma$, is inclined to form the creases. The critical loads of the surface crease and buckling behavior increase with the thickness of cylinder. When the thickness of cylinder decreases to a critical value which depends on the surface parameters, the critical load of buckling behavior is smaller than it for surface crease, and the buckling behaviors will happen before surface crease is formed. By equalling the critical pressure without surface effect in equation (29) to that in equation (30), the critical thickness can be derived to $0.677R_2$. The finite element method (FEM) is used to study the initial buckling behavior of the hollow cylinder, and the plane-strain element CPE8R in the ABAQUS finite element program is used. The critical pressures of several cylinders with different radius, thickness, and parameters of the Mooney-Rivlin materials are given by ABAQUS. The normalized critical pressures, which are independent on the radius and material parameters of the cylinder, are smaller than these from the theory model for thick cylinder as shown in Figure 6.

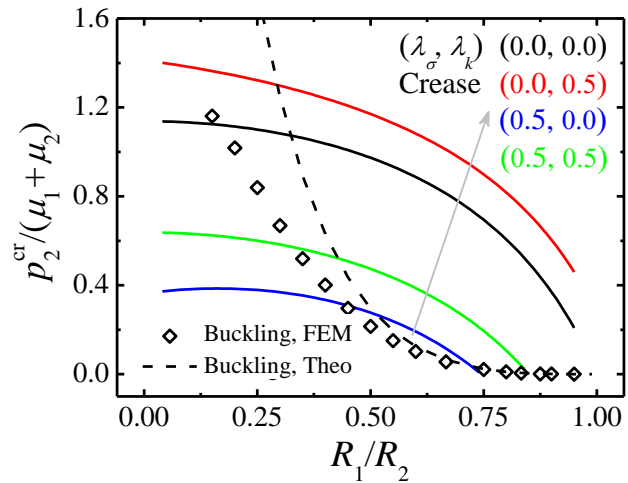

Fig. 6 The critical pressure load for the instability (i.e., creases and globe buckling) of the soft hollow cylinder for different values of surface parameters.

The outer surface of cylinder is turned to the inner surface of the deformed cylinder by geometric everting. The initial outer surface is compressive, and the corresponding compressive deformation, which increases with the thickness of cylinder, is largest along the circumference direction. When the thickness of the everted cylinder reaches to a critical value, the crease will form in the initial outer surface. By substituting equation (28) into equation (24), the ratio of the critical thickness to the initial outer radius, $\lambda_h$, can be derived from the following equation,

$$\lambda_k + \frac{\lambda_\sigma - \lambda_k}{1 + (1-\lambda_h)^{-1}} \sqrt{\beta} \left\{ \frac{1}{\sqrt{1 + \left[1-(1-\lambda_h)^2\right]\beta}} + 1 \right\}$$
$$+ \frac{\beta}{2} \left\{ \frac{(1-\lambda_h)^2}{1 + \left[1-(1-\lambda_h)^2\right]\beta} - 1 \right\} - \frac{1}{2} \ln \left\{ \frac{(1-\lambda_h)^2}{1 + \left[1-(1-\lambda_h)^2\right]\beta} \right\} = 0 \quad (31)$$

By substituting $\lambda_\sigma = 0$, $\lambda_k = 0$ into equation (31), the critical thickness of the everted cylinder without surface effect can be obtained as $0.435 R_2$, which is consistent with both the previous theoretical prediction[44] and experiment result.[40] The positive normalized residual surface stress and surface stiffness, which lead to larger deformation on the initial outer surface of the cylinder, reduce the critical thickness for the crease as shown in Fig. 7. As we know the surface parameters are not always positive, both positive and negative surface parameters can be obtained for solid with different surface properties.[17,18] For the hollow cylinder with negative surface parameters, the critical thickness is larger than it without surface effect.

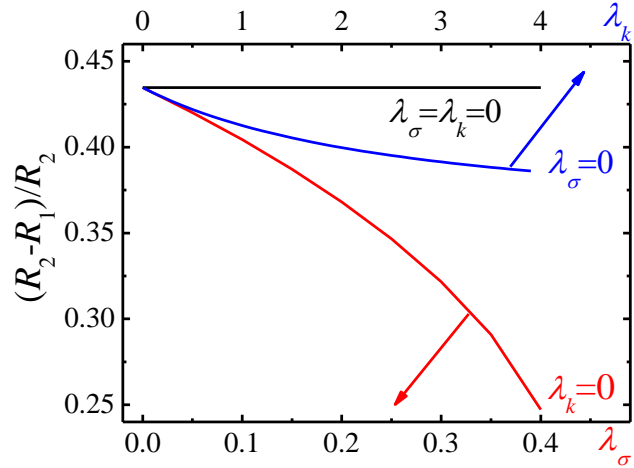

Fig. 7 The normalized critical thickness for the creases of the soft hollow cylinder under different surface parameters.

Prevent the instability of the cylinder structure has always been important but it is also a challenge, especially for the soft cylinder used in engineering environment. Through the results presented here, we can have a more clear understanding of the critical condition and the transition of instability modes. Additionally, these results also provide us an explicit instruction to design or modify the structures to prevent the instability. For example, as we know the critical condition of instability is related to the surface effects, e.g., the surface parameters, we can tune the surface parameters by employing chemical treatment to prevent the instability.

## 5. Summary and Conclusion

The surface effects on the deformation and instability of the hollow cylinder are investigated. The compressive and everted deformations of the cylinder are formulated using finite deformation theory, and are obtained analytically by employing the Moony-Rivlin constitutive model. It is found that the effects of residual surface stress and surface stiffness are different, the positive surface stiffness improved the compressed structure's overall stiffness, while the positive residual surface stress reduces it. For the soft hollow cylinder subjecting uniform pressure, the positive surface stiffness always increase the stiffness of the whole cylinder, the positive residual surface stress reduce the stiffness of the compressed cylinder, but increase the stiffness of the tensile cylinder. For the soft cylinder under the geometric everting condition, the residual surface stress and surface stiffness lead to larger deformation of the cylinder, especially thick cylinder.

Additionally, the instability of soft hollow cylinder is also significantly affected by the surface effects. For the soft cylinder under the uniform pressure, there is a transition of the instable modes from globe buckling to surface creases with increasing of the thickness of the cylinder. The phase boundary is affected by the surface effects, and is tunable by controlling the surface parameters. For the everted soft cylinder, the normalized critical thickness by the initial outer radius for the creases is decreasing with increasing surface parameters, and more significant change can be found for the variation of the residual surface stress.

Our results provide a more in-depth understanding of the effects of surface energies on the deformation and instability of soft hollow cylinders, and are expected to improve the design of soft cylinder structures in various applications. Moreover, the analysis method in the present paper can be extended to consider more complex loading conditions of soft hollow cylinder, or even the deformation and instability of soft hollow sphere.

## Acknowledgements

Authors acknowledge the supports from NSFC Grant No. 11672149 and the National Basic Research Program of China (973 Program) Grant No. 2015CB351900.

# Appendix

The nonzero components of Cauchy stress of the compressive cylinder are

$$\sigma_{rr} = -p_1 + \frac{\sigma_0}{r_1} + k_s\left(\frac{\lambda_z}{R_1} - \frac{1}{r_1}\right) + \left(\mu_1 + \mu_2\lambda_z^2\right)\frac{1}{2\lambda_z^2}\left[\frac{R^2}{r^2} - \frac{R_1^2}{r_1^2} + \lambda_z \ln\left(\frac{R^2}{R_1^2}\frac{r_1^2}{r^2}\right)\right],$$

$$\sigma_{\theta\theta} = -p_1 + \frac{\sigma_0}{r_1} + k_s\left(\frac{\lambda_z}{R_1} - \frac{1}{r_1}\right) + \left(\mu_1 + \mu_2\lambda_z^2\right)\frac{1}{2\lambda_z^2}\left[\frac{R^2}{r^2} - \frac{R_1^2}{r_1^2} + \lambda_z \ln\left(\frac{R^2}{R_1^2}\frac{r_1^2}{r^2}\right)\right]$$

$$-\left[\mu_1\left(\frac{R}{\lambda_z r}\right)^2 - \mu_2\left(\frac{\lambda_z r}{R}\right)^2\right] + \left[\mu_1\left(\frac{r}{R}\right)^2 - \mu_2\left(\frac{R}{r}\right)^2\right], \quad (A1)$$

$$\sigma_{zz} = -p_1 + \frac{\sigma_0}{r_1} + k_s\left(\frac{\lambda_z}{R_1} - \frac{1}{r_1}\right) + \left(\mu_1 + \mu_2\lambda_z^2\right)\frac{1}{2\lambda_z^2}\left[\frac{R^2}{r^2} - \frac{R_1^2}{r_1^2} + \lambda_z \ln\left(\frac{R^2}{R_1^2}\frac{r_1^2}{r^2}\right)\right]$$

$$-\left[\mu_1\left(\frac{R}{\lambda_z r}\right)^2 - \mu_2\left(\frac{\lambda_z r}{R}\right)^2\right] + \left(\mu_1\lambda_z^2 - \mu_2\lambda_z^{-2}\right).$$

The nonzero components of Cauchy stress of the everted cylinder are

$$\sigma_{rr} = \frac{\sigma_0}{r_2} + k_s\left(\frac{1}{R_2} - \frac{1}{r_2}\right) + \left(\mu_1 + \mu_2\right)\frac{1}{2}\left[\left(\frac{R^2}{r^2} - \frac{R_2^2}{r_2^2}\right) - \ln\left(\frac{R^2}{R_2^2}\frac{r_2^2}{r^2}\right)\right],$$

$$\sigma_{\theta\theta} = \frac{\sigma_0}{r_2} + k_s\left(\frac{1}{R_2} - \frac{1}{r_2}\right) + \left(\mu_1 + \mu_2\right)\frac{1}{2}\left[\left(\frac{R^2}{r^2} - \frac{R_2^2}{r_2^2}\right) - \ln\left(\frac{R^2}{R_2^2}\frac{r_2^2}{r^2}\right)\right]$$

$$-\left[\mu_1\left(\frac{R}{r}\right)^2 - \mu_2\left(\frac{r}{R}\right)^2\right] + \left[\mu_1\left(\frac{r}{R}\right)^2 - \mu_2\left(\frac{R}{r}\right)^2\right], \quad (A2)$$

$$\sigma_{zz} = \frac{\sigma_0}{r_2} + k_s\left(\frac{1}{R_2} - \frac{1}{r_2}\right) + \left(\mu_1 + \mu_2\right)\frac{1}{2}\left[\left(\frac{R^2}{r^2} - \frac{R_2^2}{r_2^2}\right) - \ln\left(\frac{R^2}{R_2^2}\frac{r_2^2}{r^2}\right)\right]$$

$$-\left[\mu_1\left(\frac{R}{r}\right)^2 - \mu_2\left(\frac{r}{R}\right)^2\right] + \left(\mu_1 - \mu_2\right).$$